\documentclass[prb,twocolumn]{revtex4}

\usepackage{graphicx}

\begin{document}



\author{T.A. Aslanyan}
 \affiliation{Institute for Physical Research, Armenian National
Academy of Sciences,  Ashtarak-2, 378410 Armenia}

\author{T. Shigenari}%
\author{K. Abe}
\affiliation{Dept. of Appl. Phys. and Chem., University of
Electro-communications, Chofu-shi, Tokyo 182, Japan}


\received{}
\begin{abstract}
We show that the method of calculations of the Debye-Waller
factors for IC structures by J.D. Axe, Phys. Rev. (1971), {\bf B
21}, 4181, is not applicable for such calculations. The set of
variables, which should be used for such calculations, is the
phase $\varphi$ and the amplitude of the IC modulation, but not
the real and imaginary parts of the complex IC modulation
function.
\end{abstract}
\maketitle ýÑÛÏ

The Debye-Waller (D-W) factors for incommensurate structures (IC)
were first calculated by Overhauser\cite{O}. From these
calculations followed that the D-W factors are tending to zero
with approaching the second order phase transition, and in some
temperature range close to the transition the IC satellite
reflections should not be observed. However, in some tens of the
discovered IC structures the IC satellites were distinctly
observed, and their intensity do not tend to zero, as it was
predicted by Overhauser. The latter became the main motivation for
the calculations by Axe\cite{Axe80}, who derived that the D-W
factors for IC structures are always close to unity, including
very close ranges near phase transition. However, in the paper by
Axe\cite{Axe80} was not pointed out the reason of such a
discrepancy between the two results, and we consider that this
problem is steel in order. We note that the result by Axe was
reproduced in the book by Bruce and Cowley\cite{BCb}, and the
result by Overhauser in the book by Krivoglaz\cite{Kriv}.

Below we compare the calculations of the D-W factors by
Axe\cite{Axe80} with those by Overhauser\cite{O} and show that the
variables used by Axe are not applicable for such calculations,
and therefore the calculations by Axe can not be treated as an
alternative to those by Overhauser. The starting point of the
discussion should be the expression for the thermodynamic
potential:
\begin{eqnarray}
\Phi=\int d{\bf R}\{\frac{\alpha}{2}\eta^2+\frac{g}{
2}(\nabla\eta)^2+ \frac{g_1}{ 2}(\nabla^2\eta)^2+\frac{b}{
4}\eta^4\}
\end{eqnarray}
where the condition for the IC transition assumes that the
coefficients $g<0$ and $g_1>0$, and the value of the IC wavevector
$k$ is given by $k^2=-g/2g_1$. The IC structure which appears
below the transition may be described using two geometrically
equivalent forms of the modulation function:
\begin{eqnarray}
\eta (R)=(\eta_0+\delta\eta )\cos ({\bf k R}+\varphi)
\end{eqnarray}
where $\eta_0$ is the equilibrium IC amplitude, and $\delta\eta$
is its fluctuation. In the equivalent form:
\begin{eqnarray}
\eta (R)=\eta_1\cos{\bf kR}+\eta_2\sin{\bf kR}
\end{eqnarray}
where $\eta_1$ and $\eta_2$ are the real and imaginary components
of the complex IC wave's amplitude $Q_{\bf k}$, i.e.,
\begin{eqnarray}
\eta (R)=Q_{\bf k}e^{i\bf kR}+Q_{-\bf k}e^{-i\bf kR}
\end{eqnarray}
and $\eta_1=(Q_{\bf k}+Q_{-\bf k})/2$, $\eta_2=i(Q_{\bf k}-Q_{-\bf
k})/2$, and besides $\eta_1^2+\eta_2^2=(\eta_0+\delta\eta)^2$,
$\eta_2/\eta_1=\tan\varphi$.

It should be emphasized that for the geometrical description of
the fluctuation-induced distortions of the IC modulation wave the
both sets of variables, namely $\delta\eta$ and $\varphi$ eq.(2)
(the IC amplitude and the IC phase), and $\eta_1$, $\eta_2$
eq.(3), are equivalent. However, calculations of the statistical
averages using these sets of variables give significantly
different results. The set of the variables $\delta\eta$ and
$\varphi$ was used in the calculations of the D-W factors for the
IC structures by Overhauser,\cite{O} Krivoglaz\cite{Kriv} and also
by the present authors\cite{we}. The variables $\eta_1$ and
$\eta_2$ were introduced for description of the IC modulation
fluctuations (phasons) by Bruce and Cowley,\cite{BC} and they were
used in the calculations of the D-W factors by Axe\cite{Axe80}. In
what follows we explain why the variables $\eta_1$ and $\eta_2$
can not be used for such calculations.

Now we introduce the variables $\eta_1$ and $\eta_2$ in the
thermodynamic potential for further calculation of the statistical
averages. We should substitute the IC modulation function eq.(3)
in the thermodynamic potential (1). In order to keep our
expressions simple we use for the IC modulation (3) the form:
\begin{eqnarray}
\eta (R)=(\bar\eta_1+\delta\eta_1)\cos{\bf kR}+
\delta\eta_2\sin{\bf kR}
\end{eqnarray}
where $\bar\eta_1$ is the equilibrium value of $\eta_1$ and
$\delta\eta_1$ is its fluctuation. In other words, we assume that
the equilibrium value of $\eta_2$ is equal to zero (i.e.,
$\bar\eta_2=0$ and $\delta\eta_2$ is its fluctuation). Therefore,
$\bar\eta_1$ is the equilibrium IC amplitude
$|\bar\eta_1|=\eta_0$, and the initial phase $\varphi_0=\arctan
(\bar\eta_2/\bar\eta_1)=0$.

Substituting (5) into eq.(1) one obtains:
\begin{eqnarray}
\Phi =\int d{\bf R}\{\frac{\tilde\alpha}{2}
\delta\eta_1^2+\frac{D}{2}(\nabla\delta\eta_1)^2+\frac{D}{
2}(\nabla\delta\eta_2)^2+ \nonumber\\
\frac{b}{32}(3\delta\eta_1^4+3\delta\eta_2^4+12\bar\eta_1\delta\eta_1^3+
6\delta\eta_1^2\delta\eta_2^2+8\bar\eta_1\delta\eta_1\delta\eta_2^2)\}
\end{eqnarray}
with $\tilde{\alpha} =-(\alpha +gk^2+g_1k^4)/2>0$,
$D=(g+6g_1k^2)/2>0$, $\bar\eta_1^2=-\tilde\alpha/b$.

The harmonic part of eq.(6) (given by the first line of eq.(6)),
in fact, was used by Axe\cite{Axe80} for calculations of the D-W
factor. As easy to see, this harmonic part is a diagonalized
quadratic form with respect to $\delta\eta_1$ and $\delta\eta_2$.
Transferring in eq.(6) to the Fourier-components (using only its
harmonic part) one can calculate in the harmonic approximation the
fluctuations:
\begin{eqnarray}
<\eta_{1k}\eta_{1-k}>=\frac{T}{V}\frac{1}{\tilde{\alpha}+Dk^2}
\nonumber\\
<\eta_{2k}\eta_{2-k}>=\frac{T}{V}\frac{1}{Dk^2}
\end{eqnarray}
Since, as it follows from this equations, the mode $\eta_2$ is a
gapless mode, it was interpreted by Axe\cite{Axe80} as a phason
mode, and the mode $\eta_1$, correspondingly, as an amplitude
mode.

Now we demonstrate that the harmonic part of eq.(6) can not be
used for such calculations in the Landau theory. Let us calculate
the fluctuation correction to the harmonic part of eq.(6) from the
inharmonic terms of the same potential\cite{pokr}. For example, a
correction from the inharmonic term
b$\bar\eta_1\delta\eta_1\delta\eta_2^2$ can be calculated after
presenting this term in the Fourier components as:
$$
b\bar\eta_1V\sum\limits_{\bf k,q}\eta_{1\bf q}\eta_{2\bf
k}\eta_{2-{\bf k}-\bf q}
$$
In the second order of expansion over this inharmonicity one will
obtain, in particular, the correction
\begin{eqnarray}
\Delta\sim \frac{V^3}{T}b^2\bar\eta_1^2|\eta_{1q}|^2\int d{\bf k}
<|\eta_{2\bf
k}|^2><|\eta_{2\bf k+q}|^2>\sim\nonumber\\
\frac{VTb^2}{D^2}\bar\eta_1^2|\eta_{1q}|^2\int d{\bf k}
\frac{1}{{\bf k}^2}\frac{1}{({\bf k+q})^2}
\end{eqnarray}
The angle brackets $<>$ in eq.(8) denote statistical averaging
with distribution function $\exp [-H_0/T]$, where $H_0$ is the
harmonic part of eq.(6) in the Fourier-form. This correction
renormalizes the coefficient at the term $\eta_{1q}\eta_{1-q}$ in
the thermodynamic potential, i.e., for $q\to 0$ it renormalizes
the coefficient $\tilde{\alpha}$ in the expansion (6). Eq.(8) can
be presented also by the graph\cite{pokr}\\[-45pt]
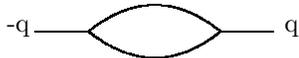
\begin{center}{\mbox{\makebox[10pt][c]{-q}\hspace{18pt}
\begin{picture}(90,40)
      \thinlines
      \put(0,0){\line(-1,0){20}}
      \put(50,0){\line(1,0){20}}
      \qbezier(0,0)(25,20)(50,0)
      \qbezier(0,0)(25,-20)(50,0)\hspace{72pt}\makebox[10pt][c]{q}
\end{picture}}}
\end{center}
where the outgoing lines correspond to the mode $\eta_1$ and the
internal coupled lines to the mode $\eta_2$.

As easy to see, for the wavevector $q\to 0$ or $q=0$ the
correction $\Delta$ diverges as
\begin{equation}
\Delta\sim \int\limits_0^{k_0} \frac{dk}{k^2}\sim \frac{1}{0}
\end{equation}
regardless of the temperature or closeness to the transition. The
latter means that the harmonic part of eq.(6) can not be used for
any calculations, since the inharmonic correction to it is not
small (it is diverging). One may expect that calculation of all
the series of the diverging graphs would cancel this divergency.
However, there is no reason to expect that the residual of such a
cancellation could be neglected.

We give an example to demonstrate that the calculations using
variables $\eta_1$, $\eta_2$ give physically inconsistent results.
Since the IC modulation function $\eta (R)$ itself is directly a
component of the crystal's electron density function it is
meaningful to discuss diffraction (of X-rays) directly by it, and
calculate the corresponding D-W factor. Let us calculate the
statistical average of the modulation function (5) using
distribution function $\exp [-H_0/T]$, where $H_0$ is the harmonic
part of eq.(6). Since $<\delta\eta_1>=0$ and $<\delta\eta_2>=0$,
the result of such averaging for eq.(5) will be $ <\eta
(R)>=\bar\eta_1\cos{\bf kR}$. In other words, statistical
averaging (in the approach adopted by Axe) does not reduce (or
affect) the amplitude $\bar\eta_1$ of the IC modulation. Since the
first order IC satellite reflection intensity is proportional to
$\bar\eta^2_1$, the latter means that the D-W factor is always
equal to unity for all temperatures. In other words, even at
temperatures $10^5$K the D-W factor appears to be equal to unity,
when it is expected to be zero, since at very high temperatures
all the Bragg-reflection intensity is expected to transfer into
the diffuse scattering.

The calculations of the D-W factors by Overhauser\cite{O}, in
fact, are based on the averaging of the modulation function $\eta
(R)$ given by eq.(2). Taking into account that the average\cite{5}
$$<\exp [i{kR}+\varphi]>=\exp [i{kR}]\exp [-<\varphi^2>/2]$$ one
can see that the average of the modulation function eq.(2) takes
the form
$$<\eta (R)>=\eta_0\cos{\bf kR}\exp [-<\varphi^2>/2],$$ where the
exponent is the D-W factor (it is less than unity) which reduces
the IC amplitude $\eta_0$ and subsequently the IC satellite
reflections. The local fluctuation of the phase $<\varphi^2>$ was
calculated by Overhauser using the thermodynamic potential as a
function of the phase $\varphi$ and amplitude $\delta\eta$ (see
also the paper by Golovko and Levanyuk\cite{Golovko}, where a more
distinct procedure is used for deriving of that potential). The
fluctuation $<\varphi^2>$ according to Overhauser diverges as
$T/\eta^2_0$ with approaching the phase transition with $\eta_0\to
0$. As a result, the D-W factor is tending to zero and attenuating
the intensity of satellite reflections. The D-W factors for IC
structures are calculated also in the paper by the present
authors\cite{we} in frame of the Landau theory using the variables
$\varphi$ and $\delta\eta$. Our result significantly differs from
that by Overhauser, and it gives not so strong divergence of the
local fluctuation $<\varphi^2>\sim T^{1/2}/\eta_0$.

In conclusion we note that though the use of the variables
$\varphi$ and $\delta\eta$ is consistent with the Landau theory
(in contrast to the variables $\delta\eta_1$ and $\delta\eta_2$),
the limits of the Landau theory applicability to the IC
transitions are not defined. One may assume that in some region
close to the transition the Landau theory is not applicable.
However, there is no reason to expect that the divergence of the
local fluctuation $<\varphi^2>$ will stop on approaching the
transition.

\end{document}